\documentclass[a4paper,10pt]{article}
\usepackage{graphicx}
\usepackage{comment}
\usepackage{amssymb}
\usepackage{amsmath}
\usepackage{nicefrac}
\usepackage{graphicx}
\usepackage{dcolumn}
\usepackage{bm}
\usepackage{comment}
\begin{document}

%
%
%

\newcommand{\bin}[2]{\left(\begin{array}{c} \!\!#1\!\! \\  \!\!#2\!\! \end{array}\right)}
\newcommand{\binc}[2]{\left[\begin{array}{c} \!\!#1\!\! \\  \!\!#2\!\! \end{array}\right]}
\newcommand{\troisj}[3]{\left(\begin{array}{ccc}#1 & #2 & #3 \\ 0 & 0 & 0 \end{array}\right)}
\newcommand{\sixj}[6]{\left\{\begin{array}{ccc}#1 & #2 & #3 \\ #4 & #5 & #6 \end{array}\right\}}
\newcommand{\ninej}[9]{\left\{\begin{array}{ccc}#1 & #2 & #3 \\ #4 & #5 & #6 \\ #7 & #8 & #9 \end{array}\right\}}
\newcommand{\ff}[3]{\mathcal{F}^{#1}(#2,#3)}
\newcommand{\gd}[3]{\mathcal{G}^{#1}(#2,#3)}

\huge

\begin{center}
Statistics of electric-quadrupole lines in atomic spectra
\end{center}

\vspace{0.5cm}

\large

\begin{center}
Jean-Christophe Pain\footnote{jean-christophe.pain@cea.fr (corresponding author)} and Franck Gilleron
\end{center}

\normalsize

\begin{center}
\it CEA, DAM, DIF, F-91297 Arpajon, France
\end{center}

\vspace{0.5cm}

\large

\begin{center}
Jacques Bauche and Claire Bauche-Arnoult
\end{center}

\normalsize

\begin{center}
\it Laboratoire Aim\'e Cotton, B\^atiment 505, Campus d'Orsay, 91405 Orsay, France
\end{center}

\vspace{0.5cm}

\begin{abstract}
In hot plasmas, a temperature of a few tens of eV is sufficient for producing highly stripped ions where multipole transitions become important. At low density, the transitions from tightly bound inner shells lead to electric-quadrupole lines which are comparable in strength with electric-dipole ones. In the present work, we propose analytical formulas for the estimation of the number of E2 lines in a transition array. Such expressions rely on statistical descriptions of electron states and $J$-levels. A generalized ``$J$-file'' sum rule for E2 lines and the strength-weighted shift and variance of the line energies of a transition array $n\ell^{N+1}\rightarrow n\ell^Nn'\ell'$ of inter-configuration E2 lines are also presented.  
\end{abstract}

\section{\label{sec1} Introduction}

In atomic spectroscopy, the transition probabilities of electric-quadrupole (E2) lines are low, as compared to electric-dipole (E1) ones. Indeed, the ratio between E2 and E1 transition probabilities is proportional to ($a_0/\lambda Z^*)^2$, where $a_0$ is the Bohr radius, $Z^*$ the effective charge of the plasma and $\lambda$ the wavelength \cite{SOBELMAN72}. This quantity is very small for most atomic transitions located in the range of UV to IR. However, they are important in astrophysics; for instance, Charro \emph{et al.} performed a careful study of intensities of E2 spectral lines in the ion Mg II along the spectral series \cite{CHARRO03,CHARRO03b}. In astrophysical plasmas of Nebulae and of the interstellar medium, the density $n_e$ of the free electrons, which are responsible for collisional excitations and de-excitations, is generally very low, and the mean time between collisions is longer than radiative E2 lifetimes, even though these are much longer than for E1 transitions. Consequently, whenever an atom gets excited by a collision onto a level, metastable or not, its only decay channel is by radiative transitions (forbidden or allowed), and forbidden transitions appear with an intensity comparable to the allowed ones \cite{KLAPISCH78}. Fournier \emph{et al.} \cite{FOURNIER96} identified by \emph{ab initio} calculations a bright $3d^8\rightarrow 3d^74s$ E2 line for Mo XVII. They also studied the ratio of $3d - 4s$ E2 to $3d - 4p$ E1 emission in Mo XVI as a function of density, and noticed that the sharp falloff in the values of this ratio for Mo XVI above certain densities provides a plasma diagnostic for planned fusion reactors (high-Z impurities breed the dilution of the fusion ``fuel'' in reactors and can have a strong impact on spatial current distributions and radiative patterns in the plasma) \cite{ISLER84,CUMMINGS90}. Atomic data of tungsten are strongly needed for identification of emission lines in future fusion reactors (ITER) where it is used as a plasma-facing material \cite{FOURNIER98}. Then, forbidden lines are of great interest for plasma diagnostics because the associated radiation intensity is very sensitive to density and temperature. Quinet gave a theoretical survey of forbidden transitions (E2 and M1) in the $4p^k$ and $4d^k$ ground configurations of ions Rh-like $W^{29+}$ to Ga-like $W^{43+}$ \cite{QUINET12}. Recently, Neu \emph{et al.} pointed out a very strong E2 spectral line in the soft X-ray region at a wavelength of 0.793 nm originating from Ni-like tungsten $W^{46+}$  \cite{NEU05}. Clementson \emph{et al.} measured this E2 line together with M3 (magnetic octupole) ground state transitions in the same ion using high-resolution crystal spectroscopy at the electron-beam ion trap (EBIT) facility in Livermore \cite{CLEMENTSON10}. EBIT light source was also used by Ralchenko \emph{et al.} \cite{RALCHENKO06,RALCHENKO07} to record spectra of Br-like $W^{39+}$ to Co-like $W^{47+}$ in the 12-20 nm region. On the other hand, in most laboratory plasmas, $n_e$ is noticeably higher. For medium-range $n_e$, the probability for collisional depopulation may be surpassed only by the probability of radiative decay through allowed transitions. In this case, quadrupole transitions are not observed, justifying their name ``forbidden''. However, they can be important in some circumstances \cite{BARSHALOM95,MIAO04,UPCRAFT06}. For instance, electric-quadrupole decays were observed  in the spectra of neon-like \cite{GAUTHIER86} and nickel-like \cite{WYART86} ions produced by laser irradiation.

As mentioned before, forbidden lines can arise from metastable levels of excited configurations \cite{COWAN81}. In highly ionized atoms having ground configurations $3p^m$ the lowest excited configuration is $3p^{m-1}3d$. For $m>1$, $p^{m-1}d$ has levels with $J'$ greater by two than the largest $J$ of $p^m$, and downward electric-dipole transitions from those levels are therefore forbidden by the dipole selection rule $\left|\Delta J\right|\leq 1$. Magnetic-dipole and electric-quadrupole transitions within the configuration Fe IX $3p^53d$ and arising from such metastable levels are responsible for several lines of the solar corona. It is interesting to mention that in Pb I, parity-forbidden transitions for electric-dipole radiation $6p^2\rightarrow 6p7p$ have been observed \cite{COWAN81}. In that case, the upper levels are not metastable, because E1 transitions to $6p7s$ are possible. In absorption, electric-quadrupole transitions of the type $s \rightarrow d$ have been extensively studied in the alkalis \cite{MOORE68,HAGAN72}.

The effective core charge (or screened nuclear charge) seen by an electron is the net charge $Z$ of the nucleus with the $N_c-1$ electrons of the ion core $Z_c=Z-N_c+1$. Along an iso-electronic sequence, the E2 radial matrix elements decrease like $Z_c^{-2}$, rather than like $Z_c^{-1}$ as do the E1 matrix elements. For inter-configuration transitions with $\Delta n\ne 0$, one has  $A_{\mathrm{E_2}}/A_{\mathrm{E_1}}\propto Z_c^2$, where $A_{\mathrm{E_1}}$ and $A_{\mathrm{E_2}}$ represent the transition probability rates for E1 and E2 lines respectively \cite{COWAN81,CHARRO01}. For inter-configuration transitions with $\Delta n=0$, the orders of magnitudes of the E1 and E2 transition probabilities vary in the same way with respect to the effective core charge $Z_c$ \cite{COWAN81}.

The detailed calculation of all the line energies and strengths in complex atomic spectra is difficult and, in some circumstances, useless. Indeed, when the densiy is sufficiently high so that the physical broadening mechanisms (\emph{e.g.} Stark shifts) are important and/or when the number of lines becomes large, the lines coalesce into broad structures. Statistical methods \cite{GINOCCHIO73,BAUCHE79,BAUCHE82,BAUCHE84,BAUCHE85,BAUCHE90,KARAZIJA91a,KARAZIJA91b,KARAZIJA92,KUCAS93,KARAZIJA95} are required because, experimentally, some quantities cannot be determined individually, but only as weighted average quantities (for instance, in emission/absorption spectra of highly ionized atoms). Moreover, explicit quantum calculations can be unappropriate, \emph{e.g.} if the hamiltonian matrix is huge. In addition, global methods can reveal physical properties hidden in a detailed treatment of levels and lines. For example, they have opened the way to the definition of the generalized $J$-file sum rule \cite{BAUCHE83} and of effective temperatures for the configuration populations \cite{BAUCHE04}.

Even for E1 lines, there exists no general compact formula for the exact number of lines in a transition array. Using group-theoretical methods, Krasnitz \cite{KRASNITZ84a,KRASNITZ84b} obtained such a compact formula only in the simple case of configurations built with non-equivalent electrons. The statistics of E1 lines was studied by Bauche and Bauche-Arnoult \cite{BAUCHE90,BAUCHE87}, but very few results were obtained as concerns the statistics of E2 lines in complex atomic spectra.  The purpose of the present work is to provide an analytical formula for the estimation of the number of E2 lines in a transition array. Such a quantity is important for opacity codes, for instance, in order to decide whether a transition array can be described statistically \cite{BAUCHE79,BAUCHE82,BAUCHE84,BAUCHE85} or requires a DLA (Detailed-Line Accounting) calculation, relying on the diagonalization of the Hamiltonian \cite{PORCHEROT11}. Like for the E1 case, a statistical description is used, because this problem does not lend itself readily to exact calculations. The first step consists in determining the statistics of the angular quantum number $J$. However, due to the fact that the quantum number $J$ is the eigenvalue of no simple operator, its mathematical study is tedious. Therefore, it is more appropriate to study the distribution of $M$, the eigenvalue (in units of $\hbar$) of the operator $J_z$ (projection along the $z-$axis of operator $\vec{J}$). The $J$ values can be obtained from the $M$ values by means of the method of Condon and Shortley \cite{CONDON35}. Any unknown distribution can be characterized by its moments. The $n^{th}$ moment of the $M$ distribution reads:

\begin{equation}
\mu_n=\frac{1}{g}\sum_{\gamma J M}\langle\gamma J M|J_z|\gamma J M\rangle^n,
\end{equation}

\noindent where the sum runs over all the $g$ states of the configuration in intermediate coupling. The distribution $P(M)$ of $M$ being symmetrical, the odd-order moments are zero. In general, it is admitted \cite{KENDALL69} that the first four moments are sufficient to capture the shape of a distribution.

In section \ref{sec2}, the second and fourth-order moments of $P(M)$ for a non-relativistic configuration are recalled. Section \ref{sec3} contains the calculation of the number of E2 lines between two configurations in the case of three different modelings of $P(M)$: Gaussian, Gram-Charlier and Generalized Gaussian. Section \ref{sec4} contains the calculation, using the aforementioned three distributions, of the number of E2 lines inside a configuration. In section \ref{sec5}, a generalized $J$-file sum rule for E2 lines is given, and in section \ref{sec6} the expressions of the strength-weighted shift and variance of a transition array of inter-configuration E2 lines are provided for $n\ell^{N+1}\rightarrow n\ell^Nn'\ell'$. Section \ref{sec7} is the conclusion.

\section{\label{sec2} Variance and kurtosis of the projection of the $J$ angular momentum}

The variance $v$ of the states of configuration $\ell^N$ is given by \cite{KUCAS93,BAUCHE87}:

\begin{equation}
v\left(\ell^N\right)=\mu_2\left(\ell^N\right)=\frac{N(4\ell+2-N)}{4\ell+1}v(\ell),
\end{equation}

\noindent where

\begin{equation}\label{varl}
v(\ell)=\frac{4\ell^2+4\ell+3}{12}.
\end{equation}

\noindent The fourth-order moment of configuration $\ell^N$ reads

\begin{equation}
\mu_4\left(\ell^N\right)=N(4\ell+2-N)[x(\ell)N(4\ell+2-N)+y(\ell)],
\end{equation}

\noindent where

\begin{equation}
x(\ell)=\frac{2\ell-1}{240(16\ell^2-1)}(40\ell^3+84\ell^2+110\ell+51)
\end{equation}

\noindent and

\begin{equation}
y(\ell)=\frac{2\ell+1}{60(16\ell^2-1)}(-16\ell^4-24\ell^3-8\ell^2+24\ell+9).
\end{equation}

\noindent For a configuration with $w$ open sub-shells $\ell_1^{N_1}\ell_2^{N_2}\ell_3^{N_3}\cdots\ell_w^{N_w}$, one has for the total variance:

\begin{equation}\label{eqjbv}
v\left(\ell_1^{N_1}\ell_2^{N_2}\ell_3^{N_3}\cdots\ell_w^{N_w}\right)=\sum_{i=1}^wv\left(\ell_i^{N_i}\right)
\end{equation}

\noindent and for the total kurtosis:

\begin{eqnarray}
\mu_4\left(\ell_1^{N_1}\ell_2^{N_2}\ell_3^{N_3}\cdots\ell_w^{N_w}\right)&=&\sum_{i=1}^w\mu_4\left(\ell_i^{N_i}\right)\nonumber\\
& &+6\sum_{i,j=1,j>i}^w\mu_2\left(\ell_i^{N_i}\right)\mu_2\left(\ell_j^{N_j}\right)
\end{eqnarray}
 
\noindent where the first sum runs over all subshells $i$ of interest, and the second sum runs over all pairs $(i,j)$ of subshells. In the following section, we will use the fourth-order reduced moment:

\begin{equation}\label{eqkurv}
\alpha_4=\frac{\mu_4}{v^2},
\end{equation}

\noindent which can also be put in the form \cite{GILLERON09}:

\begin{equation}
\alpha_4=3-\frac{1}{v^2}\sum_{i=1}^wq_i~(a_i+q_ib_i)
\end{equation}

\noindent where

\begin{equation}
a_i=\frac{(2\ell_i+1)(16\ell_i^4+24\ell_i^3+8\ell_i^2-24\ell_i-9)}{60(4\ell_i-1)},
\end{equation}

\begin{equation}
b_i=\frac{-16\ell_i^4-16\ell_i^3+88\ell_i^2+136\ell_i+3}{120(4\ell_i-1)}
\end{equation}

\noindent and

\begin{equation}
q_i=\frac{N_i(4\ell_i+2-N_i)}{4\ell_i+1}.
\end{equation}

\section{\label{sec3} Number of E2 lines between two different configurations}

The initial and final configurations are denoted $C$ and $C'$ respectively. The selection rules for E2 transitions in intermediate coupling are $\Delta J=0,\pm 1,\pm 2$ with $J+J'\ge 2$, and $\Delta \ell=0,\pm 2$ with $\ell+\ell'\ge 2$ (transitions $s\rightarrow s'$ are strictly forbidden for isolated atoms, but have been observed by means of Stark-effect-induced mixing of $n's'$ with $n''p$ states \cite{EZAWA75}). Therefore, the number $L_{E_2}$ of E2 lines between two different configurations is given by

\begin{eqnarray}\label{eqd}
L_{E_2}(C-C')&=&\sum_{J=J_{\mathrm{min}}}^{\infty}Q_C(J)\left[Q_{C'}(J)+Q_{C'}(J+1)\right.\nonumber\\
& &\left.+Q_{C'}(J-1)+Q_{C'}(J+2)+Q_{C'}(J-2)\right]\nonumber\\
& &+\epsilon'(J_{\mathrm{min}}),
\end{eqnarray}

\noindent where $Q_C(J)$ (respectively $Q_{C'}(J)$) is the number of the levels of configuration $C$ (respectively $C'$) with total angular momentum $J$ and $\epsilon(J_{\mathrm{min}})$ is a small correction, $J_{\mathrm{min}}$ being the smallest value of $J$. In many cases, $J_{\mathrm{min}}=0$ or $1/2$ according to the parity of the number of electrons. For configuration $sd$, $J_{\mathrm{min}}$=1 and for configurations involving orbitals with a large angular momentum $J$, the value of $J_{\mathrm{min}}$ can be larger. A continuous representation of Eq. (\ref{eqd}) yields 

\begin{eqnarray}\label{intbo}
L_{E_2}(C-C')&\approx&\int_{-1/2}^{\infty}Q_C(J)\left[Q_{C'}(J)+Q_{C'}(J+1)\right.\nonumber\\
& &\left.+Q_{C'}(J-1)+Q_{C'}(J+2)+Q_{C'}(J-2)\right]dJ\nonumber\\
& &+\epsilon'(J_{\mathrm{min}}),
\end{eqnarray}

\noindent where $\epsilon'(J_{\mathrm{min}})$ is a border correction. In fact, $\epsilon'(0)=\epsilon'(\frac{1}{2})=0$ are values compatible with the accuracy of the replacement of a discrete sum by an integral (see Appendix A). Using the second-order Taylor development

\begin{eqnarray}\label{eqk}
Q(J+k)&\approx&\left.Q(J)+k\frac{dQ}{dJ}\right|_{J}+\left.\frac{k^2}{2}\frac{d^2Q}{dJ^2}\right|_{J},
\end{eqnarray}

\noindent one obtains

\begin{eqnarray}
L_{E_2}(C-C')&\approx&5\int_{-1/2}^{\infty}Q_C(J)\left[Q_{C'}(J)+\frac{d^2Q_{C'}}{dJ^2}\right]dJ.
\end{eqnarray}

\noindent The method of Condon and Shortley \cite{CONDON35} enables one to express $Q(J)$ as

\begin{equation}\label{rec}
Q(J)=\sum_{M=J}^{M=J+1}(-1)^{J-M}P(M)=P(J)-P(J+1),
\end{equation}

\noindent where $P$ represents the distribution of the angular-momentum projection $M$. For a configuration $\ell_1^{N_1}\ell_2^{N_2}\ell_3^{N_3}\cdots\ell_w^{N_w}$, $P(M)$ is determined through the relation

\begin{equation}
P_{N_1,N_2,\cdots}(M)=(P_{N_1}\otimes P_{N_2}\otimes P_{N_3}\otimes\cdots\otimes P_{N_w})(M),
\end{equation}

\noindent where the distributions are convolved two at a time, which means that

\begin{equation}
(P_{N_i}\otimes P_{N_j})(M)=\sum_{M'=-\infty}^{+\infty}P_{N_i}(M')\times P_{N_j}(M-M').
\end{equation}

\noindent The total number $L_{E_2}(C-C')$ of E2 lines is invariant under the following transformations:

\vspace{5mm}

(i) The configurations $C$ and $C'$ can be interchanged.

\vspace{5mm}

(ii) In $C$ and $C'$, one can replace simultaneously all the subshells by their complementary subshells, {\emph{e.g.}: $\ell^N$ by $\ell^{4\ell+2-N}$. Therefore, the $C\rightarrow C'$ array, denoted in general $\ell_1^{N_1+1}\ell_2^{N_2}\rightarrow\ell_1^{N_1}\ell_2^{N_2+1}$ has the same number of E2 lines as its \emph{complementary} array:

\begin{equation}
\ell_1^{4\ell_1+1-N_1}\ell_2^{4\ell_2+2-N_2}\rightarrow\ell_1^{4\ell_1+2-N_1}\ell_2^{4\ell_2+1-N_2}.
\end{equation}

\vspace{5mm}

(iii) The $C\rightarrow C'$ transition array has also the same number of E2 lines as its two following \emph{semi-complementary} array

\begin{equation}
\ell_1^{4\ell_1+1-N_1}\ell_2^{N_2+1}\rightarrow\ell_1^{4\ell_1+2-N_1}\ell_2^{N_2}
\end{equation}

\noindent and

\begin{equation}
\ell_1^{N_1}\ell_2^{4\ell_2+2-N_2}\rightarrow\ell_1^{N_1+1}\ell_2^{4\ell_2+1-N_2}.
\end{equation}

\noindent The first semi-complementary array is deduced from $C\rightarrow C'$ as follows: replace the $\ell_1$ subshells by their complementaries, and exchange the $\ell_2$ subshells. For the second one, replace the $\ell_2$ subshells by their complementaries, and exchange the $\ell_1$ subshells. Complementarity ensures that the variances are unchanged, but it is not the case for semi-complementarity. However, since

\begin{eqnarray}\label{relva}
& &v\left(\ell_1^{4\ell_1+1-N_1}\right)+v\left(\ell_2^{N_2+1}\right)+v\left(\ell_1^{4\ell_1+2-N_1}\right)+v\left(\ell_2^{N_2}\right)\nonumber\\
&=&v\left(\ell_1^{N_1+1}\right)+v\left(\ell_2^{N_2}\right)+v\left(\ell_1^{N_1}\right)+v\left(\ell_2^{N_2+1}\right),
\end{eqnarray}

\noindent semi-complementarity can be ensured by replacing $v_C$ and $v_{C'}$ by their half sum $(v_C+v_{C'})/2$ \cite{BAUCHE87}. However, the kurtosis does not follow a relation similar to Eq. (\ref{relva}). It can only be ensured that the final formula for $L_{E_2}(C-C')$ be symmetrical in $C$ and $C'$, by setting $\alpha_4=(\alpha_{4,C}+\alpha_{4,C'})/2$. Following \cite{BAUCHE87}, we make the assumption that

\begin{equation}
Q(J)\approx -\left.\frac{dP}{dM}\right|_{J+1/2}.
\end{equation}

\begin{table}[!ht]
\begin{center}
\begin{tabular}{|c|c|c|c|c|}\hline\hline
Transition array & Exact & Gaussian & Gram-Charlier & GG \\\hline\hline
$d^4 \rightarrow d^3s$ & 887 & 953 & 885 & 839 \\
             &     & (+7.44 \%) & (-0.23 \%) & (-5.41 \%) \\\hline
$d^3 \rightarrow d^2g$ & 1015 & 1215 & 1092 & 1045 \\
             &      & (+19.70 \%) & (+7.56 \%) & (+2.96 \%) \\\hline
$p^3 \rightarrow p^2f$ & 110 & 124 & 116 & 80 \\
             &     & (+12.73 \%) & (+5.45 \%) & (-27.27 \%) \\\hline
$p^4 \rightarrow p^3f$ & 110 & 124 & 116 & 81 \\
             &     & (+12.73 \%) & (+5.45 \%) & (-26.36 \%) \\\hline
$d^5 \rightarrow d^4g$ & 8299 & 9550 & 8697 & 8415 \\
             &      & (+15.07 \%) & (+4.80 \%) & (+0.14 \%) \\\hline
$d^5g^3 \rightarrow d^4g^4$ & 296780266 & 299065925 & 298981701 & 293198984 \\
                  &           & (+0.77 \%) & (+0.74 \%) & (-1.21 \%) \\\hline
$d^4i^1 \rightarrow d^3g^1i^1$ & 762544 & 897641 & 783942 & 774216 \\
                     &        & (+17.72 \%) & (+2.81 \%) & (+1.54 \%) \\\hline\hline
\end{tabular}
\end{center}
\caption{\label{tab1} Number of E2 lines for different inter-configuration transitions calculated using Gaussian, fourth-order Gram-Charlier and Generalized-Gaussian modelings of $P(M)$ and compared to the exact values. An orbital $i$ corresponds to $\ell$=6.}
\end{table}

\noindent Considering the Gaussian modeling of $P(M)$:

\begin{equation}\label{gaus}
P(M)=\frac{g}{\sqrt{2\pi v}}\exp\left(-\frac{M^2}{2~v}\right),
\end{equation}

\noindent one obtains

\begin{equation}\label{gausq}
Q(J)=\frac{g}{v\sqrt{8\pi v}}(2J+1)\exp\left(-\frac{(2J+1)^2}{8~v}\right),
\end{equation}

\noindent which yields

\begin{equation}
L_{E_2}(C-C')=\frac{5g_C~g_{C'}(2v-3)}{16\sqrt{\pi}~v^{5/2}},
\end{equation}

\noindent where $g_C$ represents the degeneracy (total number of states) of configuration $C$; for instance, the degeneracy of a configuration $\ell_1^{N_1}\ell_2^{N_2}\ell_3^{N_3}\cdots\ell_w^{N_w}$ reads

\begin{equation}
g_C=\prod_{i=1}^w\bin{4\ell_i+2}{N_i}.
\end{equation}

\noindent It was also suggested in Ref. \cite{BAUCHE87} to use a fourth-order Gram-Charlier modeling of $P(M)$:

\begin{equation}\label{gram}
P(M)=\frac{g}{\sqrt{2\pi v}}\exp\left(-\frac{M^2}{2v}\right)\left[1+\frac{(\alpha_4-3)}{24}\left(3-6\frac{M^2}{v}+\frac{M^4}{v^2}\right)\right].
\end{equation}

\noindent In that case, one finds

\begin{eqnarray}
Q(J)&=&\frac{g}{v\sqrt{8\pi v}}\left[(2J+1)+\frac{(\alpha_4-3)}{24}\left(\vphantom{\frac{(2J+1)^5}{16v^2}}15(2J+1)\right.\right.\nonumber\\
& &\left.\left.-10\frac{(2J+1)^3}{4~v}+\frac{(2J+1)^5}{16~v^2}\right)\right]\exp\left(-\frac{(2J+1)^2}{8~v}\right),
\end{eqnarray}

\noindent which yields

\begin{eqnarray}
L_{E_2}(C-C')&=&\frac{1}{16384\sqrt{\pi}~v^{5/2}}\left[5g_C~g_{C'}(-6747+2018~v\right.\nonumber\\
& &\left.+5~\alpha_4(938-124~v+21~\alpha_4(-11+2v)))\right].
\end{eqnarray}

\noindent Finally, considering the Generalized-Gaussian modeling of $P(M)$:

\begin{equation}\label{gg}
P(M)=\frac{g}{\sqrt{v}}~\frac{e^{-\left|\frac{M}{\lambda\sqrt{v}}\right|^\nu}}{
2\lambda~\Gamma\left(1+\frac{1}{\nu}\right)}
\;\;\;\mathrm{with}\;\;\;
\lambda=\sqrt{\frac{\Gamma\left(\frac{1}{\nu}\right)}{\Gamma\left(\frac{3}{\nu}\right)}},
\end{equation}

\noindent where $\nu$ is a positive real number, and $\Gamma(x)$ is the ordinary gamma function, one has

\begin{equation}\label{qjgg}
Q(J)=\frac{g_C\nu}{2\lambda^2v~\Gamma\left(1+\frac{1}{\nu}\right)}\left(\frac{2J+1}{2\lambda\sqrt{v}}\right)^{\nu-1}\exp\left(-\left[\frac{2J+1}{2\lambda\sqrt{v}}\right]^{\nu-1}\right)
\end{equation}

\noindent and the resulting number of E2 lines is

\begin{eqnarray}
L_{E_2}(C-C')&=&\frac{5~g_C~g_{C'}\nu^3 2^{\frac{1}{\nu}}}{64\lambda^3v^{3/2}~\Gamma\left(\frac{1}{\nu}\right)^2}\left[-\frac{4^{\frac{1}{\nu}}(\nu-1)(2\nu-1)}{\lambda^2v}~\Gamma\left(2-\frac{3}{\nu}\right)\right.\nonumber\\
& &+\left.4~\Gamma\left(2-\frac{1}{\nu}\right)\right].
\end{eqnarray}

\noindent As can be seen in table \ref{tab1}, the fourth-order Gram-Charlier expansion series provides a better agreement in most of the cases, except for $d^4i^1 \rightarrow d^3g^1i^1$ and $d^5 \rightarrow d^4g^1$. This is due to the fact that, when a high-$\ell$ electron is involved, the shape of the distribution $P(M)$ is not quasi-Gaussian anymore, but exhibits a plateau \cite{GILLERON09}, which cannot be modeled by Gram-Charlier expansion. On the other hand, the Generalized Gaussian can depict such situations (the ``door function'' corresponds to an exponent equal to $\nu=9/8$). Furthermore, the second-order Taylor-series expansion (see Eq. (\ref{eqk})) is not as precise as in the E1 case, since the selection rules here imply that the maximum value of $k$ is 2 (against 1 in the E1 case).

\section{\label{sec4} Number of E2 lines in the intra-configuration case ($C=C'$)}

\begin{table}[!ht]
\begin{center}
\begin{tabular}{|c|c|c|c|c|}\hline\hline
Configuration & Exact & Gaussian & Gram-Charlier & GG \\\hline\hline
$d^6$ & 358 & 380 & 352 & 335 \\
      &     & (+6.15 \%) & (-1.68 \%) & (-6.42 \%) \\\hline
$d^5p$ & 15383 & 16499 & 15446 & 14967 \\
       &       & (+7.25 \%) & (+0.41 \%) & (-2.70 \%) \\\hline
$p^2f$ & 353 & 402 & 358 & 322 \\
       &     & (+13.88 \%) & (+1.42 \%) & (-8.78 \%) \\\hline
$p^3f$ & 611 & 696 & 622 & 567 \\
       &     & (+13.91 \%) & (+1.80 \%) & (-7.20 \%) \\\hline
$g^7$ & 903622 & 968412 & 909392 & 892418 \\
      &        & (+7.17 \%) & (+0.64 \%) & (-1.24 \%) \\\hline
$f^7p^2$ & 5437574 & 5839649 & 5481760 & 5373233 \\
         &         & (+7.39 \%) & (+0.81 \%) & (-1.18 \%) \\\hline
$f^7d^2$ & 40832855 & 43424047 & 41104770 & 40379223 \\
         &          & (+6.35 \%) & (+0.67 \%) & (-1.11 \%) \\\hline
$f^3p^2$ & 97615 & 107207 & 98683 & 96137 \\
         &       & (+9.83 \%) & (+1.09 \%) & (-1.51 \%) \\\hline
$f^3d^2$ & 690870 & 746580 & 696669 & 681553 \\
         &        & (+0.81 \%) & (+0.84 \%) & (-1.35 \%) \\\hline
$d^2l$ & 2376 & 2545 & 2064 & 2365 \\
       &      & (+7.11 \%) & (-13.13 \%) & (-0.46 \%) \\\hline
$p^3i$ & 616 & 655 & 529 & 581 \\
       &     & (+6.33 \%) & (-14.12 \%) & (-5.68 \%) \\\hline\hline
\end{tabular}
\end{center}
\caption{\label{tab2} Number of E2 lines inside different configurations calculated using Gaussian, fourth-order Gram-Charlier and Generalized-Gaussian modelings of $P(M)$ and compared to the exact values. An orbital $l$ corresponds to $\ell$=8.}
\end{table}

The number of E2 lines inside a configuration can be estimated as

\begin{equation}
L_{E_2}\approx\int_{-1/2}^{\infty}Q(J)\left[\frac{1}{2}\left(Q(J)-1\right)+Q(J+1)+Q(J+2)\right]dJ.
\end{equation}

\noindent Using the second-order Taylor development presented in Eq. (\ref{eqk}), one finds the approximate expression

\begin{eqnarray}
L_{E_2}&\approx&\frac{1}{2}\int_{-1/2}^{\infty}Q(J)\left[5\left\{Q(J)+\frac{d^2Q}{dJ^2}\right\}+6\frac{dQ}{dJ}-1\right]dJ.
\end{eqnarray}

\noindent A Gaussian expression of $P(M)$ (see Eq. (\ref{gaus})) leads to

\begin{equation}
L_{E_2}=\frac{g_C(-8\sqrt{2}~v_C^2+5~g_C(2~v_C-3))}{32\sqrt{\pi}~v_C^{5/2}},
\end{equation}

\noindent and a fourth-order Gram-Charlier expression of $P(M)$ (see Eq. (\ref{gram})) to

\begin{eqnarray}
L_{E_2}&=&\frac{-2048~g_C(\alpha_{4,C}+5)v_C^2}{32768\sqrt{2\pi}~v_C^{5/2}}\nonumber\\
&&+\frac{5\sqrt{2}~g_C^2}{32768\sqrt{2\pi}~v_C^{5/2}}\left\{-6747+2018~v_C+5~\alpha_{4,C}[938-124~v_C\right.\nonumber\\
& &\left.+21~\alpha_{4,C}(2~v_C-11)]\right\}.
\end{eqnarray}

\noindent Using a Generalized-Gaussian expression of $P(M)$ (see Eq. (\ref{gg})), one obtains

\begin{eqnarray}
L_{E_2}&=&\frac{5~g_C^2\nu^3 2^{1/\nu}}{128\lambda^3v_C^{3/2}~\Gamma\left(\frac{1}{\nu}\right)^2}\left[-\frac{4^{1/\nu}(\nu-1)(2\nu-1)}{\lambda^2v_C}~\Gamma\left(2-\frac{3}{\nu}\right)\right.\nonumber\\
& &\left.+4~\Gamma\left(2-\frac{1}{\nu}\right)\vphantom{\frac{1}{2}}\right]-\frac{\nu g_C}{4\lambda\sqrt{v_C}~\Gamma\left(\frac{1}{\nu}\right)}.
\end{eqnarray}

\noindent Here also the fourth-order Gram-Charlier expansion series provides a better agreement in most of the cases, except for configurations with high-$\ell$ electrons (see table \ref{tab2}). The results are closer to the exact values in the intra-configuration case than in the inter-configuration case. This is due to the fact that the number of lines in the intra-configuration case is evaluated with the exact variance and the exact kurtosis of the distribution of $M$ (in the inter-configuration case, they were obtained as the arithmetic average over the initial and final configurations). 

It is worth mentioning that global methods were also used for the study of statistical properties of Auger amplitudes and rates \cite{KYNIENE02} and the dispersion of gyromagnetic ratios in complex spectra \cite{ROSENZWEIG61}. In appendix C, we give, following the work of Kynien\.{e} \emph{et al.} \cite{KYNIENE02}, an approximate expression for the number of Auger amplitudes with the three different modelings of $P(M)$.

\section{\label{sec5} A generalized $J$-file sum rule for E2 lines}

Using the second-quantization technique \cite{JUDD67}, Bauche \emph{et al.} \cite{BAUCHE83} established a linear relationship between the angular coefficient of the exchange Slater integral $G^1(n\ell,n'\ell-1)$ and the $J$-file sums of the E1 line strengths defined by Condon and Shortley \cite{CONDON35}. In the case of E2 lines, a $J$-file sum rule can be also obtained, following the same procedure as for E1 lines. The sum of the strengths of all E2 lines starting from a level $\gamma J$ of the upper configuration in the $\ell^N\ell'^{N'+1}\rightarrow \ell^{N+1}\ell'^{N'}$ array is given by: 
 
\begin{eqnarray}
S_{E2}\left[\left(\ell^N\ell'^{N'+1}\right)\gamma J - \ell^{N+1}\ell'^{N'}\right]&=&(2J+1)\left[\frac{(N'+1)}{(2\ell'+1)}\langle\ell||C^{(2)}||\ell'\rangle^2\right.\nonumber\\
& &\left.+C(G^2;\gamma J)\vphantom{\frac{(N'+1)}{(2\ell'+1)}}\right]\left[I(n\ell,n'\ell')\right]^2\nonumber\\
\end{eqnarray}

\noindent with

\begin{equation}
I(n\ell,n'\ell')=\int_0^{\infty}R_{n\ell}(r)r^2R_{n'\ell'}(r)dr
\end{equation}

\noindent and

\begin{equation}
\langle\ell||C^{(2)}||\ell'\rangle=(-1)^{\ell}\sqrt{(2\ell+1)(2\ell'+1)}\troisj{\ell}{2}{\ell'}.
\end{equation}

\noindent The quantity $C(G^2;\gamma J)$ represents the coefficient of the $G^2(n\ell,n'\ell')$ Slater integral in the electrostatic energy of the level $\gamma J$. In the present case, one has, if $|\ell'-\ell|=2$ and $\ell_>=\max(\ell,\ell')$:

\begin{equation}\label{c21}
\langle\ell||C^{(2)}||\ell'\rangle^2=\frac{3\ell_>(\ell_>-1)}{2(2\ell_>-1)},
\end{equation}

\noindent and if $\ell=\ell'$:

\begin{equation}\label{c22}
\langle\ell||C^{(2)}||\ell'\rangle^2=\frac{\ell(\ell+1)(2\ell+1)}{(2\ell-1)(2\ell+3)}.
\end{equation}

\noindent These results remain unchanged if passive subshells are added to both configurations. It is interesting to mention that the total strength of transition array $C\rightarrow C'$ with $C=n\ell^{N+1}n'\ell'^{N'}n''\ell''^{N''}\cdots$ and $C'=n\ell^Nn'\ell'^{N'+1}n''\ell^{N''}\cdots$ reads

\begin{eqnarray}
T(C\rightarrow C')&=&2\frac{(N+1)(4\ell'+2-N')}{(4\ell+2)(4\ell'+2)}g_C\langle\ell||C^{(2)}||\ell'\rangle^2\left[I(n\ell,n'\ell')\right]^2,\nonumber\\
\end{eqnarray}

\noindent where $g_C$ represents the degeneracy of configuration C, \emph{i.e.}

\begin{equation}
g_C=\bin{4\ell+2}{N+1}\bin{4\ell'+2}{N'}\bin{4\ell''+2}{N''}\cdots.
\end{equation}

\section{\label{sec6} Shift and variance of a transition array $n\ell^{N+1}\rightarrow n\ell^Nn'\ell'$ of E2 lines}

To our knowledge \cite{BAUCHE88}, Harrison and Johnson \cite{HARRISON31} are the first who introduced the term ``transition array'' for the entire bunch of lines resulting from transitions between two configurations. The first experimental spectrum showing transition arrays was published by Edl\'en in 1947 \cite{EDLEN47}: it concerned transition elements in the XUV range and the spectra were generated by low-inductance discharge lamps. Since then, such arrays have been observed in a very large variety of spectra (see for instance \cite{KLAPISCH82,SVENDSEN94,PALMERI96,KUCAS05,LOISEL09}). Although, in tokamak applications, E2 lines appear usually as isolated (only a few distinct lines are visible), it is also possible to estimate the global properties (strength-weighted moments) of a transition array of E2 lines. In the present section, we consider the case of inter-configuration E2 lines for the specific transition array $n\ell^{N+1}\rightarrow n\ell^Nn'\ell'$ with $\ell'-\ell=0,\pm 2$ and $\ell+\ell'\ge 2$. The moments of the distribution of the line energies can be written as:

\begin{equation}
\mathcal{M}_n=\frac{\sum_{a,b}\left[\langle b|H|b\rangle-\langle a|H|a\rangle\right]^n|\langle a|Q|b\rangle|^2}{\sum_{a,b}|\langle a|Q|b\rangle|^2}
\end{equation}

\noindent where $a$ and $b$ run over all the exact eigenstates of the Hamiltonian in configurations $C$ and $C'$, respectively, and $Q$ is the $z$ component of the quadrupole transition operator. The first two moments were calculated by Bauche-Arnoult \emph{et al.} \cite{BAUCHE79} using the second-quantization techniques of Judd \cite{JUDD67}. The most useful quantities are the mean energy $\mathcal{M}_1$ and the variance $\sigma^2=\mathcal{M}_2-\mathcal{M}_1^2$ related to the spectral width of the UTA. For the mean energy of the array, it was found in Ref. \cite{BAUCHE79} that

\begin{equation}
\mathcal{M}_1=E_{C'}-E_C+\delta E,
\end{equation}

\noindent where $E_C$ and $E_{C'}$ are the average energies of the initial and final configurations respectively. There exists a shift between the weighted average energy of a transition array and the difference of average energies of the initial and final configurations. The shift $\delta E$ stems from the weighting factor $\langle a|Q|b\rangle|^2$ and is non zero for transition arrays $n\ell^{N+1}\rightarrow n\ell^Nn'\ell'$. It was shown in Ref. \cite{BAUCHE79} that $\sigma^2$ can be written as

\begin{equation}
\sigma^2=\sum_ic_i\left[\sum_{k,k'}d_i(k,k',\ell,\ell',\cdots)\times e_i(n\ell,n'\ell',\cdots)\vphantom{\sum_{k,k'}}\right],
\end{equation}

\noindent where $c_i$ are numerical coefficients depending on the number of equivalent electrons $N$, $d_i(k,k',\ell,\ell',\cdots)$ are combinations of $3nj$ ($n$=1,2 and 3) symbols independent of $N$ and $e_i(n\ell,n'\ell',\cdots)$ are products of Slater integrals of ranks $k$ and $k'$. In the following, $F_{C}^k$ and $F_{C'}^k$ represent the direct Slater integrals in configurations $C$ and $C'$, respectively, and $G_{C'}^k$ the exchange Slater integral of $C'$. As in Ref. \cite{BAUCHE79}, we define the quantities:

\begin{eqnarray}
x&=&N(N+1)(4\ell-N)(4\ell-N+1)=(N+1)(4\ell-N)w,\nonumber\\
y&=&N(N-1)(4\ell-N+1)(4\ell-N+2)=(N-1)(4\ell-N+2)w,\nonumber\\
z&=&N(N-1)(4\ell-N)(4\ell-N+1)=(N-1)(4\ell-N)w,\nonumber\\
u&=&N(4\ell-N)(4\ell-N+1)=(4\ell-N)w,\nonumber\\
v&=&N(N-1)(4\ell-N+1)=(N-1)w,\nonumber\\
w&=&N(4\ell-N+1).
\end{eqnarray}

\noindent In the following, in agreement with the convention of Ref. \cite{BAUCHE79}, indices $n$ and $n'$ are omitted in the Slater integrals. The shift is given by:

\begin{equation}
\delta E=N\frac{(2\ell+1)(2\ell'+1)}{(4\ell+1)}\left(\sum_{k\ne 0}f_kF_{C'}^k(\ell\ell')+\sum_kg_kG_{C'}^k(\ell\ell')\right)
\end{equation}

\noindent with

\begin{equation}
f_k=\troisj{\ell}{k}{\ell}\troisj{\ell'}{k}{\ell'}\sixj{\ell}{k}{\ell}{\ell'}{2}{\ell'},
\end{equation}

\noindent and

\begin{equation}
g_k=\troisj{\ell}{k}{\ell'}^2\left(\frac{2}{5}\delta_{k,2}-\frac{1}{2(2\ell+1)(2\ell'+1)}\right),
\end{equation}

\noindent where $\delta_{i,j}$ represents a Kronecker's symbol. The variance can be written as 

\begin{equation}
\sigma^2=\sum_{i=1}^7H_i. 
\end{equation}

\noindent As compared to Ref. \cite{BAUCHE79}, the term $H_1$ remains unchanged:

\begin{eqnarray}
H_1&=&\sum_{k\ne 0}\sum_{k'\ne 0}\left(\frac{2\delta_{k,k'}}{(2k+1)}-\frac{1}{(2\ell+1)(4\ell+1)}\right.\nonumber\\
& &\left.-(-1)^{k+k'}\sixj{\ell}{\ell}{k}{\ell}{\ell}{k'}\right)\frac{(2\ell+1)^3}{(4\ell-1)8\ell(4\ell+1)}\nonumber\\
& &\times\troisj{\ell}{k}{\ell}^2\troisj{\ell}{k'}{\ell}^2\nonumber\\
& &\times\left[xF_C^k(\ell\ell)F_C^{k'}(\ell\ell)+yF_{C'}^k(\ell\ell)F_{C'}^{k'}(\ell\ell)-2zF_C^k(\ell\ell)F_{C'}^{k'}(\ell\ell)\right],
\end{eqnarray}

\noindent and one has

\begin{eqnarray}
H_2&=&\sum_{k\ne 0}\sum_{k'\ne 0}\left(\frac{2(-1)^k}{(2k+1)}\delta_{k,k'}\sixj{\ell'}{\ell'}{k}{\ell}{\ell}{2}\right.\nonumber\\
& &-(-1)^k\sixj{\ell}{\ell}{k'}{\ell}{\ell}{k}\sixj{\ell}{\ell}{k'}{\ell'}{\ell'}{2}\nonumber\\
& &\left.-\frac{1}{(2\ell+1)(4\ell+1)}\sixj{\ell}{\ell}{k'}{\ell'}{\ell'}{2}\right)\frac{(2\ell+1)^3(2\ell'+1)}{(4\ell-1)2\ell(4\ell+1)}\nonumber\\
& &\times\troisj{\ell}{k}{\ell}^2\troisj{\ell'}{k'}{\ell'}\troisj{\ell}{k'}{\ell}\nonumber\\
& &\times\left[uF_C^k(\ell\ell)F_{C'}^{k'}(\ell\ell')+vF_{C'}^k(\ell\ell)F_{C'}^{k'}(\ell\ell')\right],
\end{eqnarray}

\begin{eqnarray}
H_3&=&\sum_{k\ne 0}\sum_{k'}\left[-2\sixj{k}{k'}{2}{\ell'}{\ell}{\ell}+\sixj{\ell'}{\ell'}{k}{\ell}{\ell}{2}\sixj{\ell'}{\ell'}{k}{\ell}{\ell}{k'}\right.\nonumber\\
& &-\left.\frac{1}{(2\ell+1)(4\ell+1)}\left(\frac{2}{5}\delta_{k',2}-\frac{1}{(2\ell'+1)}\right)\right]\frac{(2\ell+1)^3(2\ell'+1)}{(4\ell-1)2\ell(4\ell+1)}\nonumber\\
& &\times\troisj{\ell}{k'}{\ell'}^2\troisj{\ell}{k}{\ell}^2\nonumber\\
& &\times\left[uF_C^k(\ell\ell)G_{C'}^{k'}(\ell\ell')+vF_{C'}^k(\ell\ell)G_{C'}^{k'}(\ell\ell')\right],
\end{eqnarray}

\begin{eqnarray}
H_4&=&\sum_{k\ne 0}\sum_{k'\ne 0}\left(\frac{2\delta_{k,k'}}{(2\ell'+1)(2k+1)}-\ninej{\ell}{\ell}{k}{\ell}{2}{\ell'}{k'}{\ell'}{\ell'}\right.\nonumber\\
& &-\left.\frac{1}{(4\ell+1)}\sixj{\ell}{\ell'}{2}{\ell'}{\ell}{k}\sixj{\ell}{\ell'}{2}{\ell'}{\ell}{k'}\vphantom{\ninej{\ell}{\ell}{k}{\ell}{2}{\ell'}{k'}{\ell'}{\ell'}}\right)\frac{(2\ell+1)^2(2\ell'+1)^2}{4\ell(4\ell+1)}\nonumber\\
& &\times\troisj{\ell}{k}{\ell}\troisj{\ell}{k'}{\ell}\troisj{\ell'}{k}{\ell'}\troisj{\ell'}{k'}{\ell'}\nonumber\\
& &\times wF_{C'}^k(\ell\ell')F_{C'}^{k'}(\ell\ell'),
\end{eqnarray}

\begin{eqnarray}
H_5&=&\sum_{k}\sum_{k'}\left[\frac{2\delta_{k,k'}}{(2\ell'+1)(2k+1)}-\ninej{\ell}{\ell'}{k}{\ell'}{2}{\ell}{k'}{\ell}{\ell'}\right.\nonumber\\
& &-\left.\frac{1}{(4\ell+1)}\left(\frac{2}{5}\delta_{k,2}-\frac{1}{(2\ell'+1)}\right)\left(\frac{2}{5}\delta_{k',2}-\frac{1}{2\ell'+1}\right)\vphantom{\ninej{\ell}{\ell'}{k}{\ell'}{2}{\ell}{k'}{\ell}{\ell'}}\right]\nonumber\\
& &\times\frac{(2\ell+1)^2(2\ell'+1)^2}{4\ell(4\ell+1)}\troisj{\ell}{k}{\ell'}^2\troisj{\ell}{k'}{\ell'}^2\nonumber\\
& &\times wG_{C'}^k(\ell\ell')G_{C'}^{k'}(\ell\ell'),
\end{eqnarray}

\noindent and the last term (for the electrostatic part):

\begin{eqnarray}
H_6&=&\sum_{k\ne 0}\sum_{k'}\left[\frac{(-1)^k}{(2\ell'+1)}\sixj{\ell'}{\ell'}{k}{\ell}{\ell}{k'}\right.\nonumber\\
& &-2(-1)^k\sixj{k}{k'}{2}{\ell}{\ell'}{\ell'}\sixj{k}{k'}{2}{\ell'}{\ell}{\ell}-\frac{1}{(4\ell+1)}\nonumber\\
& &\left.\times\sixj{\ell}{\ell'}{2}{\ell'}{\ell}{k}\left(\frac{2}{5}\delta_{k',2}-\frac{1}{(2\ell'+1)}\right)\right]\frac{(2\ell+1)^2(2\ell'+1)^2}{2\ell(4\ell+1)}\nonumber\\
& &\times\troisj{\ell}{k}{\ell}\troisj{\ell'}{k}{\ell'}\troisj{\ell}{k'}{\ell'}^2\nonumber\\
& &\times wF_{C'}^k(\ell\ell')G_{C'}^{k'}(\ell\ell').
\end{eqnarray}

\noindent The spin-orbit contribution is the same as for the E1 case \cite{BAUCHE79}:

\begin{eqnarray}
H_7&=&(N+1)(4\ell-N+1)\frac{\ell(\ell+1)}{4(4\ell+1)}\zeta_C^2(n\ell)+\nonumber\\
& &+N(4\ell-N+2)\frac{\ell(\ell+1)}{4(4\ell+1)}\zeta_{C'}^2(n\ell)+\frac{\ell'(\ell'+1)}{4}\zeta_{C'}^2(n'\ell')\nonumber\\
& &-N(4\ell-N+1)\frac{\ell(\ell+1)}{2(4\ell+1)}\zeta_C(n\ell)\zeta_{C'}(n\ell)\nonumber\\
& &-(4\ell-N+1)\frac{\ell(\ell+1)+\ell'(\ell'+1)-2}{4(4\ell+1)}\zeta_C(n\ell)\zeta_{C'}(n'\ell')\nonumber\\
& &-N\frac{\ell(\ell+1)+\ell'(\ell'+1)-2}{4(4\ell+1)}\zeta_{C'}(n\ell)\zeta_{C'}(n'\ell'),\nonumber\\
\end{eqnarray}

\noindent where $\zeta_C(n\ell)$ and $\zeta_{C'}(n\ell)$ are the spin-orbit integrals in the respective configurations $C$ and $C'$. Higher moments can also be calculated \cite{BAUCHE84,KARAZIJA91a,KARAZIJA91b,KARAZIJA92,JONAUSKAS07}, which requires to use another modeling function than the Gaussian. The Generalized Gaussian method \cite{GILLERON08} can be applied, but the Normal Inverse Gaussian was shown to provide a better depiction of the array when the first four moments are known \cite{PAIN09,PAIN10}.

The range of energy of the levels of the upper configuration $C'$ responsible for the preferential emission constitutes an ``emissive zone'' \cite{BAUCHE83}. In the same way as for an E1 UTA, the shift and width of the emissive zone of an E2 UTA are deduced from those of the complete UTA by restricting them to the radial parameters related to the upper configuration of the array.

\section{\label{sec7} Conclusion}

We propose analytical expressions for the number of electric-quadrupole (E2) lines both in the inter- and intra-configuration cases. The resulting formulas are based on three different modelings of the distribution of the angular-momentum projection $M$ (Gaussian, fourth-order Gram-Charlier and Generalized Gaussian). The Gram-Charlier modeling gives satisfactory results but, for high-$\ell$ electrons, the Generalized-Gaussian distribution is more accurate (as for E1 lines, see Ref. \cite{GILLERON09}). The results are better in the intra- than in the inter-configuration case, which can be explained by the fact that the number of lines between two different configurations is evaluated with averaged parameters (variance and kurtosis of the distribution of the angular-momentum projection $M$). We also provide the expression of the generalized $J$-file sum rule giving the total strength of the lines arising from a given level of the upper configuration, together with the formula for the average strength-weighted position and variance of E2 lines. The next step will consist in investigating the statistics of the amplitudes and strengths of E2 lines. 

\section{Appendix A: Boundary effect on the number of lines of an array}

At very small values of $J$, we can linearize the exponential in Eq. (\ref{gausq}), \emph{i.e.}

\begin{equation}\label{lin1}
Q_C(J)=K_C\times(2J+1)
\end{equation}

\noindent and

\begin{equation}\label{lin2}
Q_{C'}(J)=K_{C'}\times(2J+1),
\end{equation}

\noindent where $K_C$ and $K_{C'}$ do not depend on $J$.

\vspace{5mm}

(i) Inter-configuration case:

\vspace{5mm}

In the inter-configuration case, the error committed in considering that $\epsilon'(\frac{1}{2})=0$ in Eq. (\ref{intbo}) is equal to

\begin{equation}
\Delta_{\epsilon,\mathrm{inter}}=5\int_{-1/2}^0Q_C(J)Q_{C'}(J)dJ
\end{equation}

\noindent Using the linearized expressions (\ref{lin1}) and (\ref{lin2}), one obtains

\begin{equation}
\Delta_{\epsilon,\mathrm{inter}}\approx\frac{5}{6}(K_C+K_{C'}).
\end{equation}

\vspace{5mm}

(ii) Intra-configuration case:

\vspace{5mm}

In the intra-configuration case, the error committed in considering that $\epsilon'(\frac{1}{2})=0$ in Eq. (\ref{intbo}) is equal to

\begin{equation}
\Delta_{\epsilon,\mathrm{intra}}=3\int_{-1/2}^0Q_C(J)Q_{C'}(J)dJ
\end{equation}

\noindent Using the linearized expressions (\ref{lin1}) and (\ref{lin2}), one obtains

\begin{equation}
\Delta_{\epsilon,\mathrm{intra}}\approx\frac{K_C+K_{C'}}{2}.
\end{equation}

\section{Appendix B: Estimation of $J_{\mathrm{min}}$}

Usually, $J_{\mathrm{min}}$ is equal to 0 (for integer values) or to $1/2$ (for half-integer values). However, in some circumstances \cite{GILLERON09} (for instance in the case of configurations containing an electron in an orbital with a high angular momentum $\ell$), the distribution $P(M)$ exhibits a plateau, for which we showed that the Generalized Gaussian in a good aproximation \cite{GILLERON09}. In that case, $J_{\mathrm{min}}$ differs from 0 and $1/2$. An estimation of $J_{\mathrm{min}}$ can be obtained through the relation:

\begin{equation}
Q(J)\ge \frac{1}{2}.
\end{equation}

\noindent Using the expression of $Q(J)$ given in Eq. (\ref{qjgg}), we find

\begin{equation}
J_{\mathrm{min}}\approx\lambda\sqrt{v}\left(\frac{1-\nu}{\nu}\right)W\left[\frac{\nu}{1-\nu}\Lambda^{\frac{\nu}{\nu-1}}\right]^{1/\nu}-\frac{1}{2},
\end{equation}
 
\noindent with 

\begin{equation}
\Lambda=\frac{\lambda^2~v_C}{\nu^2g_C}\Gamma\left(\frac{1}{\nu}\right),
\end{equation}

\noindent and $x\mapsto W[x]$ represents Lambert's function, solution of $x~e^x=y$. The function $W$ can be expanded as \cite{LAMBERT1758,COMTET70,JEFFREY95,HASSANI05}:

\begin{equation}\label{dev}
W(x)=\ln x -\ln(\ln x)+\sum_{k=0}^{\infty}\sum_{m=1}^{\infty}c_{km}\frac{[\ln(\ln x)]^m}{(\ln x)^{k+m}}
\end{equation}

\noindent where 

\begin{equation}
c_{km}=\frac{(-1)^k}{m!}S[k+m,k+1], 
\end{equation}

\noindent $S[p,q]$ being Stirling number of the first kind \cite{COMTET74,MITRINOVIC60}, also denoted $S_p^{(q)}$, $s(p,q)$ or $\binc{p}{q}$. Stirling numbers can be obtained by recursion relations \cite{COMTET70}, and an explicit expression was provided by Karanicoloff \cite{KARANICOLOFF61}. However, since we are only interested in an approximate formula, the first two terms $\ln x -\ln(\ln x)$ are sufficient.

\section{Appendix C: Number of Auger amplitudes}

Auto-ionization from a state that involves a hole in an inner subshell of the core is known as the Auger \cite{AUGER25} effect, and the ejected electron is called an Auger electron. The term ``auto-ionization'' (applied to levels produced by excitation of loosely bound electrons) was coined by Shenstone \cite{SHENSTONE31}. In the non-relativistic approximation the amplitude of Auger transitions is equal to the reduced matrix element of the Coulomb interaction operator $H_c$:

\begin{equation}
\langle C\gamma J||H_c||C'\gamma'J'\epsilon\ell jJ\rangle=\sqrt{2J+1}~\langle C\gamma J|H_c|C'\gamma'J'\epsilon\ell jJ\rangle,
\end{equation}

\noindent where $C$ is the configuration of an atom, $J$ is the quantum number of total angular momentum, $\gamma$ denotes all the additional quantum numbers and $\epsilon$ is the energy of the Auger electron. In intermediale coupling, the number of Auger amplitudes for a given channel $\epsilon_{\ell}$ or $\epsilon_{\ell j}$ can be obtained as the number of reduced matrix elements of the scalar operator acting between two configurations:

\begin{equation}
N_{\mathrm{Auger}}(C-C')\approx\int_0^{\infty}Q_C(J)Q_{C''}(J)dJ,
\end{equation}

\noindent where $C''$ denotes $C'\epsilon \ell$ and $C'$ is the final configuration of the ion. In the configuration $C''$, the contribution of the Auger electron to the variance and kurtosis of the distribution $P(M)$ is equal to the one of an electron of the discrete spectrum $v(\epsilon\ell)=v(\ell)$, where $v(\ell)$ is defined in Eq. (\ref{varl}) and

\begin{equation}
\mu_4(\epsilon\ell)=\frac{1}{240}(2\ell+1)(48\ell^4+96\ell^3+152\ell^2+104\ell+15).
\end{equation}

\noindent It was shown by Kynien\.{e} {\emph{et al.} \cite{KYNIENE02} that the Auger transition 

\begin{equation}
\ell_1^{4\ell_1+1}\ell_2^{N_2}\ell_3^{N_3}\rightarrow\ell_1^{4\ell_1+2}\ell_2^{N_2-1}\ell_3^{N_3-1}\epsilon\ell
\end{equation}

\noindent does not change under the replacements $N_2\rightarrow 4\ell_2+4-N_2$ and $N_3\rightarrow 4\ell_3+4-N_3$. In the same way, the transition

\begin{equation}
\ell_1^{4\ell_1+1}\ell_2^{N_2}\rightarrow\ell_1^{4\ell_1+2}\ell_2^{N_2-2}\epsilon\ell.
\end{equation}

\noindent does not change under the replacement $N_2\rightarrow 4\ell_2+4-N_2$. This is a consequence of complementarity, and can be explained by the fact that the recoupling does not change the number of matrix elements \cite{KYNIENE02}. Here also, since 

\begin{eqnarray}\label{relvauger}
& &v\left(\ell_1^{4\ell_1+1}\right)+v\left(\ell_2^{N_2}\right)+v\left(\ell_1^{4\ell_1+2}\right)+v\left(\ell_2^{N_2-2}\right)+v(\epsilon\ell)\nonumber\\
&=&v\left(\ell_1^{4\ell_1+1}\right)+v\left(\ell_2^{4\ell_2+4-N_2}\right)+v\left(\ell_1^{4\ell_1+2}\right)+v\left(\ell_2^{4\ell_2+2-N_2}\right)+v(\epsilon\ell),\nonumber\\
& &
\end{eqnarray}

\noindent we set $v=(v_C+v_{C'})/2$ and, in order to obtain an expression symmetrical in C and $C'$, $\alpha_4=(\alpha_{4,C}+\alpha_{4,C'})/2$. Using the Gaussian expression of $P(M)$ (see Eq. (\ref{gaus}), Kynien\.{e} \emph{et al.} obtained \cite{KYNIENE02}:

\begin{equation}
N_{\mathrm{Auger}}(C-C')\approx\frac{g_C~g_{C'}}{8\sqrt{\pi}~v^{3/2}}.
\end{equation}

\noindent The Gram-Charlier modeling of $P(M)$ (see Eq. (\ref{gram}) leads to

\begin{equation}
N_{\mathrm{Auger}}(C-C')\approx g_C~g_{C'}\frac{1009+5~\alpha_4(-62+21~\alpha_4)}{8192\sqrt{\pi}~v^{3/2}},
\end{equation}

\noindent and using a Generalized-Gaussian approximation of $P(M)$, we find

\begin{equation}
N_{\mathrm{Auger}}(C-C')\approx g_C~g_{C'}\frac{1}{2^{2+1/\nu}\lambda\Gamma\left(1+\frac{1}{\nu}\right)\sqrt{v}}.
\end{equation}

\end{document}